\documentclass[reqno]{article}
\newcommand{\ds}{\displaystyle}
\newcommand{\dsf}{\ds\frac}

\newcommand{\beq}{\begin{equation}}
\newcommand{\eeq}{\end{equation}}

\usepackage{ae} 
\usepackage[T1]{fontenc}
\usepackage[ansinew]{inputenc}
\usepackage{amsmath}
\usepackage{graphicx}
\usepackage{color}
\usepackage[colorlinks]{hyperref}
\usepackage{multicol}
\begin{document}
\footnotesize

\begin{center}
\bf  On the flux jumps in the flux creep regime of type - II
superconductors
\end{center}

\begin{center}
N. A. Taylanov
\end{center}

\begin{center}
\emph{National University of Uzbekistan}
\end{center}

\begin{center}
{\bf Abstract}
\end{center}

\begin{center}
\mbox{\parbox{13cm}{\footnotesize The spatial and temporal
evolution of small perturbations of the temperature and
electromagnetic field for the superconducting slab placed in a
parallel magnetic field in the regime of thermally activated flux
creep is studied. On the basis of theoretical analysis of thermal
diffusion and Maxwell's equations we find the onset field $B_j$ of
the flux-jump instability field and its dependence on the external
magnetic-field sweep rate $\dot{B_e}$.}}
\end{center}


\vskip 1cm

\begin{multicols}{2}{\vskip 0.5 cm
\begin{center}
{\bf Introduction}
\end{center}

As we know that the conventional II type superconductor are
characterized by a high value of the critical current and upper
critical field, and is therefore widely used for various
technological applications. One of the main topic at the technical
application of superconductors is their stability with respect to
flux jumps [1-3]. The flux jumps or avalanches are associated with
a sudden puncture of magnetic flux into the volume of the sample,
and in turn, an increase of temperature and the decrease of
critical current density. The jump phenomena have been observed in
conventional hard superconductors [1-6], as well as in
high-temperature superconductors, recently [7, 8]. The critical
state stability against flux jumps in hard and composite
superconductors has been discussed in a number of theoretical and
experimental papers [1-6]. The general concept of the
thermomagnetic instabilities in type-II superconductors was
developed in literature [4, 5]. The dynamics of small thermal and
electromagnetic perturbations, whose development leads to the flux
jump, have been investigated theoretically in detail by Mints and
Rakhmanov [5]. The authors have found the stability criterion for
the flux jumps in the framework of adiabatic and dynamic
approximations in the viscous flux flow regime of type-II
superconductors. Conventionally, thermomagnetic instabilities were
interpreted in terms of thermal runaway triggered by local energy
dissipation in the sample [5]. According to this theory, any local
instability causes a small temperature rise, the critical current
is decreased and magnetic flux moves much easily under the Lorentz
force. The additional flux movement dissipates more energy further
increasing temperature. This positive feedback loop may lead to a
flux jumps in the superconductor sample.

Theoretical investigations of small thermal and electromagnetic
perturbations in a various regimes with a various current-voltage
characteristics is one of key problems of electrodynamics of
superconductors. Moreover, there is a special interest to this
problem in the flux creep regime of type-II superconductors, since
all superconducting devices used for a large scale applications
operate under this regime. In our previous work, the dynamics of
small thermal and electromagnetic perturbations has been studied
in the flux flow regime, where voltage current-current
characteristics of hard superconductor is described by linear
dependence of $\vec j(\vec E)$ at sufficiently large values of
electric field $\vec E$ [9]. In the region of weak electric fields
the current-voltage characteristics $\vec j(\vec E)$ of
superconductors is highly nonlinear due to thermally activated
dissipative flux motion. In the flux creep regime a differential
conductivity $\sigma$ strongly depends on the electric field $E$.
The nonlinear conductivity $\sigma(E)$ significantly affects the
dynamics of thermal and electromagnetic processes in
superconductors. In particular, it results in the dependence of
the flux-jump field $B_j$ on variation of external parameters, in
particular on the magnetic field sweep rate $\dot{B_e}$. A
theoretical analyze of the flux jumping in the flux creep regime,
where the current-voltage characteristics of a sample is a
nonlinear have been carried out recently by Mints [10] and by
Mints and Brandt [11]. However, a careful study the dynamics of
the thermal and electromagnetic perturbations in the regime weak
electric field with nonlinear current-voltage characteristics
associated with flux creep is still lacking.

\vskip 0.5 cm
\begin{center}
{\bf Objectives}
\end{center}

We report the results of theoretical simulations of the spatial
and temporal evolution of small perturbations of the temperature
and electromagnetic field for the superconducting slab placed in a
parallel magnetic field in the regime of thermally activated flux
creep. On the basis of theoretical analysis of thermal diffusion
and Maxwell equations we find the onset field $B_j$ of the
flux-jump instability field and its dependence on the external
magnetic-field sweep rate $\dot{B_e}$. It is assumed that the
magnetic diffusion is slower than the thermal diffusion.

\vskip 0.5 cm
\begin{center}
{\bf 1. Formulation of the problem}
\end{center}

Mathematical problem of theoretical study the dynamics of thermal
and electromagnetic perturbations in a superconductor sample in
the flux creep regime can be formulated on the basis of a system
nonlinear diffusion-like equations for the thermal and
electromagnetic field perturbations with account nonlinear
relationship between the field and current in superconductor
sample. Bean [1] has proposed the critical state model which is
successfully used to describe magnetic properties of type II
superconductors. According to this model, the distribution of the
magnetic flux density $\vec B$ and the transport current density
$\vec j$ inside a superconductor is given by a solution of the
equation

\begin{equation}
rot\vec B=\vec j.
\end{equation}

When the penetrated magnetic flux changes with time, an electric
field $\vec E$ is generated inside the sample according to
Faraday's law
\begin{equation}
rot\vec E=\frac{d\vec B}{dt}.
\end{equation}

The temperature distribution in superconductor is governed by the
heat conduction diffusion equation

\begin{equation}
\nu (T)\dsf{dT}{dt}=\nabla[\kappa(T)\nabla T]+\vec j\vec E,
\end{equation}

Here $\nu=\nu(T)$ and $\kappa=\kappa(T)$ are the specific heat and
thermal conductivity, respectively. The above equations should be
supplemented by a current-voltage characteristics of
superconductors, which has the form
$$
j=j_{c}(T,B)+j(E).
$$
In order to obtain analytical results of a set Eqs. (1)-(3), we
suggest that $j_c$ is independent on magnetic field induction $B$
and use the Bean critical state model $j_c=j_c(B_e, T)$ [1], where
$B_e$ is the external applied magnetic field induction. We assume
that the dependence of the critical current density on temperature
of the sample is linear [9]. For the sake of simplifying of the
calculations, we perform our calculations on the assumption of
negligibly small heating $(T-T_0\ll T_c-T_0)$ and assume that the
temperature profile is a constant within the across sample and
thermal conductivity $\kappa$ and heat capacity $\nu$ are
independent on the temperature profile; where $T_0$ and $T_c$ are
the equilibrium and critical temperatures of the sample,
respectively, [9].  We shall study the problem in the framework of
a macroscopic approach, in which all lengths scales are larger
than the flux-line spacing; thus, the superconductor is considered
as an uniform medium.

The system of differential equations (1)-(3) should be
supplemented by a current- voltage curve $j=j(E)$. In the flux
creep regime the current-voltage characteristics of type II
conventional superconductors is highly nonlinear due to thermally
activated dissipative flux motion [12, 13]. For the logarithmic
current dependence of the potential barrier $U(j)$, proposed by
[14] the dependence $j(E)$ has the form

\begin{equation}
j=j_c\left[\dsf{E}{E_c}\right]^{1/n}.
\end{equation}
where the constant parameter n depends on the pinning regimes and
can vary widely for various types of superconductors. In the case
$n=1$ the power-law relation (4) reduces to Ohm's law, describing
the normal or flux-flow regime [15]. For infinitely large $n$, the
equation describes the Bean critical state model $j=j_c$ [1]. When
$1<n<\infty$, the equation (4) describes nonlinear flux creep. In
this case the differential conductivity $\sigma$ is determined by
the following expression

\begin{equation}
\sigma=\dsf{d\vec j}{d\vec E}=\dsf{j_c}{nE_b}.
\end{equation}
According to relation (5) the differential conductivity decreases
with the increasing of  the background electric field $E$, and
strongly depends on the external magnetic field sweep rate
$E_b\sim \dot B_ex$. Therefore the stability criterion also
strongly depends on the differential conductivity $\sigma$. For
the typical values of $j_1=10^8 A/cm^2$, $E_b=10^{-2} V/cm$ we
obtain $\sigma=10^{10} 1/\Omega cm$. It follows from this
estimation [11] that the differential conductivity $\sigma$ which
determines the dynamics of thermomagnetic instability for is high
enough. We assume, for simplicity, that the value of n temperature
and magnetic-field independent.

\vskip 0.5 cm
\begin{center}
{\bf 2. Basic equations}
\end{center}

Let us formulate a differential equations governing the dynamics
of small temperature and electromagnetic field perturbation in a
superconductor sample. We study the evolution of the thermal and
electromagnetic penetration process in a simple geometry -
superconducting semi-infinitive sample $x\geq 0$. We assume that
the external magnetic field induction $B_e$ is parallel to the
z-axis and the magnetic field sweep rate $\dot{B_e}$ is constant.
When the magnetic field with the flux density $B_e$ is applied in
the direction of the z-axis, the transport current $\delta j(x,
t)$ and the electric field $\delta E(x, t)$ are induced inside the
slab along the y-axis. For this geometry the spatial and temporal
evolution of small thermal $\delta T(x, t)$ and electromagnetic
field $\delta E(x, t)$ perturbations

\begin{equation}
\delta T=\Theta(x)\exp[\gamma t],
\end{equation}
\begin{equation}
\delta E=\epsilon(x)\exp[\gamma t].
\end{equation}
is described by Maxwell equations coupled to the thermal diffusion
equation

\begin{equation}
\nu\gamma\Theta=\kappa\dsf{d^2\Theta}{dx^2}+j_c\epsilon,
\end{equation}
\begin{equation}
\dsf{d^2\epsilon}{dx^2}=\dsf{4\pi}{c^2}\gamma\left[\dsf{j_c}{nE_b}\epsilon-\dsf{j_c}{T_c-T_0}\Theta\right].
\end{equation}
where $\gamma$ is the eigenvalue of the problem to be determined.
It is clear that the rate $\gamma$ characterizes the time
development of the instability. In the case when Re$\gamma\geq 0$,
small thermal and electromagnetic perturbations increase and the
stability margin corresponds to the case when $\gamma$=0. It
should be noted that the nonlinear diffusion-type equations (8)
and (9), totally determine the problem of the space-time
distribution of the temperature and electromagnetic field profiles
in the flux creep regime with a nonlinear current-voltage
characteristics in the semi-infinite sample.

\vskip 1 cm
\begin{center}
{\bf 3. Dispersion relation}
\end{center}

Let us derive the dispersion equation to determine the eigenvalue
problem. As we know [5] that a nature of the flux jumps depends on
the competition between diffusive and dissipative processes
through the dimensionless parameter
$$
\tau=\dsf{4\pi\sigma\kappa}{c^2\nu}=\dsf{D_t}{D_m}.
$$
where $D_t=\kappa/\nu$ is the thermal diffusivity and
$D_m=c^2/4\pi\sigma_f$ the magnetic diffusivity coefficients,
respectively. Therefore the flux instability criterion is
determined mainly by the relation of the magnetic $D_m$ and
thermal $D_t$ diffusion coefficients. As we have mentioned above,
the differential conductivity $\sigma(E)$, which determines the
dynamics of the instability is high in the flux creep regime and
the parameter $\tau$ is high enough, also. It is clear that this
picture for the flux jumps corresponds to the limiting case
$\tau\gg 1$. Consequently, it can be assumed that the initial
rapid heating stage of a flux jump takes place on the background
of a "frozen-in" magnetic flux.  Therefore, under this dynamic
approximation, we obtain from (9) the relation between electric
field $\epsilon(x, t)$ and temperature $\Theta(x, t)$
perturbations in the following form

\begin{equation}
\dsf{j_c}{nE}\epsilon-\dsf{j_c}{T_c-T_0}\Theta=0.
\end{equation}
We notice that the last relation between $\epsilon(x, t)$ and
temperature $\Theta(x, t)$ has been derived in the assumption that
the decrease of the critical current density $j_c$ resulting from
a temperature perturbation $\Theta(x,t)$ compensates with increase
of the resistive current density $j_r$ resulting from an electric
field perturbation $\epsilon(x, t)$, so the total current density
remains constant [5]. Upon substituting the expression (10) into
the equation (8) and excluding the variable $\epsilon(x, t)$ one
can get the differential equation for the distribution of thermal
perturbation, which can be conveniently presented in the following
dimensionless form

\begin{equation}
\dsf{d^2\Theta}{d\rho^2}- \rho\Theta=0.
\end{equation}
Here we introduced the following dimensionless variables
$$
\rho=\dsf{\gamma-z}{r},\quad \dsf{1}{r}=\left[n\dsf{a
L^2}{\kappa}E_{L}\right]^{1/3}, \quad E_L\simeq \dot{B_e}L, \quad
a=\dsf{j_c}{T_c-T_0}, \quad z=\dsf{x}{L}.
$$
Here $L=\dsf{cB_e}{4\pi j_c}$ is the magnetic field penetration
depth. Thus, the condition of existence of a non-trivial solutions
of Eq. (11) allows to define the spectrum of eigenvalues of
$\gamma$ and the instability threshold, accordingly. The equation
(11) has an exact solution in terms of Airy functions given as the
following form

\begin{equation}
\Theta (\rho)=c_1(s)Ai(\rho)+c_2(s)Bi(\rho).
\end{equation}
where $Ai(\rho)$ and $Bi(\rho)$ are the Airy functions. Here
constants of integration $c_1$ and $c_2$ are determined from the
thermal boundary conditions. Substituting the last solution (12)
into the thermal boundary conditions (see, for example [9]) we
find that $c_2=0$ and $\Theta (\rho)=c_1(s)Ai(\rho)$. Applying the
second boundary condition $\Theta(1)=0$ we get an equation to
determine the eigenvalues of the problem

$$
\textcolor{black}{J_{2/3}(a_n)=J_{-2/3}(a_n)}.
$$
where $a_n$ are the zeros of the Bessel function and growing with
increasing n; For example, for n=1 the stability criterion is
presented as
\begin{equation}
a_1=r^{2/3}\gamma.
\end{equation}
Using the value for the magnetic field penetration depth, we can
easily obtain from (13) an expression for the threshold magnetic
field $B_j$ at which the branching instability occurs
\begin{equation}
B_j=\dsf{4\pi j_c}{c}\sqrt{\dsf{\kappa}{anE_{L}}}.
\end{equation}
Let us now estimate the threshold field for a typical values of
parameters $j_c\simeq 10^9 A/m^2$, $T_c-T_0\simeq$ 10 K,
$\kappa\simeq 10^{-1} W/K m$, n=10. The background electric field
$E_L\simeq \dot{B_e} L$, induced by the magnetic-field variation
$\dot{B_e}\simeq 10^{-2}\div 10^{-3}$ T/s is of the order of
$E_L=10^{-4}\div 10^{-5}$ V/m for the value of $L=0.01$ m. We can
easily estimate that the threshold field has the value $B_j\asymp
1\div 3 T$.

\vskip 0.5cm
\begin{center}
{\bf 4. Discussion}
\end{center}

Experimentally, the background electric field is created by the
sweeping rate of applied magnetic field $\dot{B_e}$. As can be
seen from the relation (14) the threshold field $B_j$ is decreased
with the increasing of background electric field. It is noticeable
that the dependence of the flux-jump field $B_j$ on the sweeping
rate $\dot{B_e}$ of the applied magnetic field have been verified
by a numerous experiments [8, 16-18]. An intensive numerical
analysis on the sweep rate dependence of the threshold field has
been performed recently in [19]. Recent magnetization measurements
[8, 16] have shown that the value of the threshold field $B_j$
decreases as the sweep rate $\dot{B_e}$ increases. A theoretical
investigations on the dependence of threshold field on the varying
external magnetic field has been performed in detail recently by
Mints [10]. Within the framework of the flux jump instability
theory [4, 5] a rapid variation of the applied magnetic field acts
as the instability-driving perturbation, and that threshold field
$B_j$ should decrease with increasing the sweeping rate
$\dot{B_e}$ [8]. The numerical studies [19] have demonstrated that
the flux jumps takes place when the sweep rate $\dot{B_e}$
increases up to a certain value, where the number of jumps
increases with the sweep rate $\dot{B_e}$. As the sweep rate
further increases, these simulation results show that the
flux-jump field decreases and approaches a saturation value, which
is fairly close to the experimental value of about $1.2$ T/s [8].
However, as has been mentioned in [16], an experimental
investigations on the dependence of the threshold field for the
flux jump field $B_j$ on the external magnetic field sweep rate
$\dot{B_e}$ is very little. Experimentally, can be observed a
complex behavior of dependence of the threshold field $B_j$ on the
sweep rate $\dot{B_e}$. The results of experiments of Ref. [16]
demonstrated that $B_j$ is independent of the sweep rate in a
defined range of temperatures. Thus, the near independence of
$B_j$ on the sweeping rate remains to be explained. In some
conventional superconductors both the independence of $B_j$ on the
sweeping rate [3, 4] and its growth at a high sweeping rate [20]
were detected. It has been suggested [4] that a nonuniform heating
may be responsible for such an effect. However, theoretical
understanding of the thermomagnetic instabilities at such
conditions is still lacking. Note, however, that some details of
the local field behavior depend indeed on the sweeping rate, as,
for example, the number and amplitude of the jumps. Gerber at. al.
[18] have demonstrated that at low values of the sweep rates the
number of flux jumps decreases as sweep rate increases. At still
high sweep rates the amplitude of flux jumps becomes independent
of the sweep rate and saturates to the limit with further
increasing sweep rate.

\begin{center}
\includegraphics[width=2.5583in]{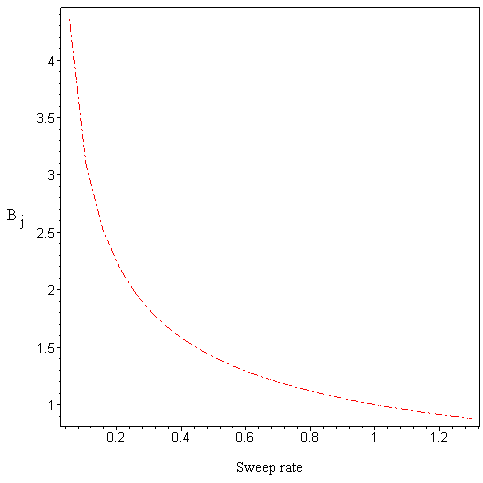}
\end{center}
\begin{center}
Fig.1  The sweep rate dependence of the field of the flux jump
field.
\end{center}

In Fig.1 we have demonstrated the dependence of the threshold
field $B_j$ on the external magnetic field sweep rate. As can be
seen, the value of $B_j$ decreases as the sweep rate increases. As
the sweep rate increases the value of $B_{j}$ decreases and it
tends to saturate at high sweep rates [8]. Magnetic field
dependence of the critical current density only slows down the
decrease of the field $B_j$ with increasing external magnetic
field sweep rate [10, 11]. We note that for the case Kim-Anderson
model [12], the absolute value of the exponent in the power
formula decreases from 1/2 to 1/3, so $B_j\sim \dot{B_e}^{-1/3}$
[10].

Let us qualitatively estimate the temperature dependence of the
first flux jump field $B_{j}$(T). In order to compare the last
formula with experimentally determined first flux jump field, one
must know the temperature dependencies of the critical current
density and the specific heat of the sample. As has been mentioned
in literature [5, 16] a quantitative estimation of $j_c$(T) from
the available experimental and theoretical models is difficult
because of the uncertainty in the values of the critical current
$j_c$ at a given field and temperature. To determine $j_c(T)$,
different approaches have been taken. Empirically, the critical
current density $j_c$(T) can be presented in the form
$$
j_c=j_0\left[1-t^n\right]^m.
$$
where $1<m<2$, $t=T/T_c$. The different exponents n=1 and 2 refer
to the most common cases discussed in the literature, where the
critical current exhibits a linear and a quadratic dependence on
$T/T_c$. There is experimental evidence [21], which indicates that
the temperature dependence of the critical current density
approximately linear at low temperatures. However, at higher
temperatures, where flux creep effects are dominant, the
temperature dependence of the critical current density can be
presented as

$$
j_c(T)=j_c(0)\left(1-t^2\right)^2.
$$
commonly accepted in literature [6]. Assuming that the thermal
conductivity is a linear function of temperature, we can easily
obtain an expression for the temperature dependence of the
threshold field $B_{j}$(T) (Fig 2.)

$$
B_{j}(T)\approx\sqrt{t\left(1-t^2\right)^2}.
$$

\begin{center}
\includegraphics[width=2.5583in]{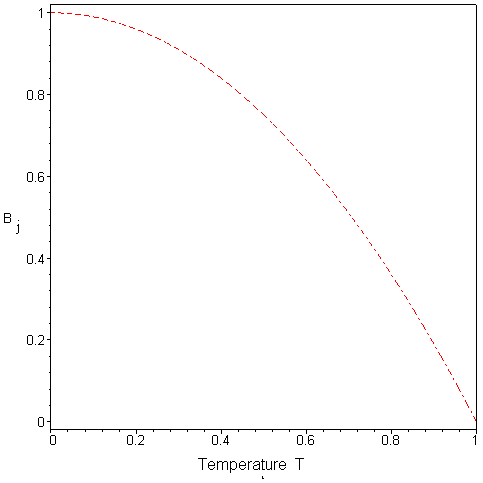}
\end{center}
\begin{center}

Fig.2. The temperature dependence of the flux jump field
\end{center}

\begin{center}
{\bf Conclusion}
\end{center}

We have performed a theoretical study of dynamics of small thermal
and electromagnetic perturbations in type-II superconductors in
the flux creep regime with a nonlinear current-voltage
characteristics. For this purpose, the space-time evolution of
temperature and electric field was calculated using the heat
diffusion equation, coupled with Maxwell’s equations and material
law, assuming that the applied magnetic field is directed parallel
to the surface of the sample. We found the threshold field $B_j$
for the occurrence of thermomagnetic instability assuming that the
heat flux diffusion is considerable faster than the magnetic flux
diffusion in superconductor. The obtained stability criterion for
the thermomagnetic flux jumps demonstrates the extremely high
sensitivity of the threshold field $B_j$ on the values of the
critical current density $j_c$, thermal conductivity $\kappa$, and
external magnetic field sweep rate $\dot{B_e}$. Thus, the
stability condition (14) for the thermomagnetic flux jumps
directly reflects the magnetic sweep rate $\dot{B_e}$ dependence
on the threshold field $B_j$. It follows from the criterion that
the value of the threshold field $B_j$ is inversely proportional
to the square root of the magnetic-field sweeping rate
$\dot{B_e}$. Therefore, with the increase of sweeping rate
$\dot{B_e}$ the threshold field $B_j$ decreases. Finally, we have
discussed the temperature dependence of the threshold field,
briefly.

\vskip 0.5cm
\begin{center}
{\bf Acknowledgements}
\end{center}
This study was supported by the NATO Reintegration Fellowship
Grant and Volkswagen Foundation Grant. Part of the computational
work herein was carried on in the Condensed Matter Physics at the
Abdus Salam International Centre for Theoretical Physics.

\vskip 0.5cm
\begin{center}
{\bf References}
\end{center}

\begin{enumerate}
\item C. P. Bean, Phys. Rev. Lett., 8, 250, 1962; Rev. Mod. Phys.,
36, 31, 1964.

\item P. S. Swartz and S. P. Bean, J. Appl. Phys., 39, 4991, 1968.

\item S. L. Wipf, Cryogenics, 31, 936, 1961.

\item  R. G. Mints and A.L. Rakhmanov, Rev. Mod. Phys., 53, 551,
1981.

\item R. G. Mints and A.L. Rakhmanov, Instabilities in
superconductors, Moscow, Nauka, 362, 1984.

\item  A. M. Campbell and J. E. Evetts, Critical Currents in
Superconductors, (Taylor and Francis, London, 1972);  Moscow,
1975.

\item L. Legrand, I. Rosenman, Ch. Simon, and G. Collin, Physica
C, 211, 239, 1993.

\item A. Nabialek,  M. Niewczas, Physica C, 436, 43, 2006.

\item N. A. Tayalanov and A. Elmuradov, Technical Physics, 11, 48,
2003.

\item R. G. Mints, Phys. Rev., B 53, 12311, 1996.

\item R. G. Mints and E. H. Brandt, Phys. Rev., B 54, 12421, 1996.

\item P. W. Anderson , Y. B. Kim  Rev. Mod. Phys., 36. 1964.

\item P. W. Anderson,  Phys. Rev. Lett.,  309, 317, 1962.

\item E. Zeldov, N. M. Amer, G. Koren, A. Gupta, R. J. Gambino,
and M. W. McElfresh, Phys. Rev. Lett., 62, 3093, 1989.

\item P. H. Kes, J. Aarts, J. van der Berg, C. J. van der Beek,
and J.A. Mydosh, Supercond. Sci. Technol., 1, 242, 1989.

\item  P. Esquinazi, A. Setzer, D. Fuchs, Y. Kopelevich, E.
Zeldov, C. Assmann, Phys. Rev. B, 60, 12454, 1999.

\item  D. Stamopoulos, A. Speliotis and D. Niarchos,
cond-mat-0410570v1, 2004

\item  A. Gerber, Z. Tarnawski, and J. J. M. Franse, Physica C
209, 147, 1993.

\item  You-He Zhou and Xiaobin Yang, Phys. Rev. B, 74, 054507,
2006.

\item Y. B. Kim, C. F. Hempstead, and A. R. Strnad, Phys. Rev.
Letters, 9, 306, 1962.

\item A. El Bindari and M. M. Litvak, J. Appl. Phys., 34, 2913,
1963.

\end{enumerate}

}\end{multicols}

\end{document}